\documentclass[preprint,aps,prd,showpacs,nofootinbib,superscriptaddress,tightenlines]{revtex4}

\usepackage{amsfonts}
\usepackage{mathrsfs}
\usepackage{graphicx}
\usepackage{amsmath}
\usepackage{amssymb}
\usepackage{bm}
\usepackage{bbm}
\usepackage{multirow}
\usepackage{alltt}
\usepackage{fancyvrb}
\usepackage[usenames]{color}
\usepackage[T1]{fontenc}

\def\nn{\nonumber}
\def\gs#1{#1\!\!\!/}

\def\PA#1{\left( #1 \right)}
\def\PB#1{\left[ #1 \right]}
\def\PC#1{\left\{ #1 \right\}}

\def\mw{M_W}
\def\mbc{M_{B_c}}
\def\xb{\bar{x}}

\def\exchange{\begin{array}{c}e_c \to e_b \\ x_0 \leftrightarrow \bar{x}_0 \end{array}}
\def\xchange{\begin{array}{c}e_c \to e_b \\ x \leftrightarrow \bar{x} \end{array}}

\def\OMIT#1{}

\newcommand{\beq}{\begin{equation}}
\newcommand{\eeq}{\end{equation}}
\newcommand{\bqa}{\begin{eqnarray}}
\newcommand{\eqa}{\end{eqnarray}}

\begin{document}
%\title{Refactorization for the Amplitude of ${\rm W}$ Radiative Decay into $B_c$ Through
%Incorporating NRQCD and Light-Cone Approaches}
%%%%%%%%%%%%%%%%%%%%%%%%%%%%%%%%%%%%%%%%%%%%%%%%%%%%%%%%%%%%%%%%%%%%%%%%%%%%%%
\title{Optimized predictions for $W \to B_c + \gamma$ by combining light-cone and NRQCD approaches}

\author{Feng Feng\footnote{F.Feng@outlook.com}}
\affiliation{China University of Mining and Technology, Beijing 100083, China\vspace{0.2cm}}
\affiliation{Institute of High Energy Physics, Chinese Academy of
Sciences, Beijing 100049, China\vspace{0.2cm}}
\affiliation{Center for High Energy Physics, Peking University, Beijing 100871, China\vspace{0.2cm}}

\author{Yu Jia\footnote{jiay@ihep.ac.cn}}
\affiliation{Institute of High Energy Physics, Chinese Academy of
Sciences, Beijing 100049, China\vspace{0.2cm}}
\affiliation{Center for High Energy Physics, Peking University, Beijing 100871, China\vspace{0.2cm}}
\affiliation{School of Physics, University of Chinese Academy of Sciences,
Beijing 100049, China\vspace{0.2cm}}
%\affiliation{Center
%for High Energy Physics, Peking University, Beijing 100871,
%China\vspace{0.2cm}}

\author{Wen-Long Sang\footnote{wlsang@ihep.ac.cn}}
 \affiliation{School of Physical Science and Technology, Southwest University, Chongqing 400700, China\vspace{0.2cm}}

\date{\today}
%%%%%%%%%%%%%%%%%%%%%%%%%%%%%%%%%%%%%%%%%%%%%%%%%%%%%%%%%%%%%%%%%%%%%%%%%%%%%%
\begin{abstract}
%%%%%%%%%%%%%%%%%%%%%%%%%%%%%%%%%%%%%%%%%%%%%%%%%%%%%%%%%%%%%%%%%%%%%%%%%%%%%%
The rare process of $W$ radiative decay into the $B_c$ meson can serve as an interesting play ground
for testing two influential perturbative QCD approaches: non-relativistic quantum chromodynamics (NRQCD)
and ligh-cone (LC) factorization. Since the $B_c$ meson is a genuine heavy quarkonium which
is composed of two different species of heavy quarks, it is natural to employ NRQCD factorization to
tackle this exclusive quarkonium production process; on the other hand, since the mass of the $W$ boson
is much greater than that of the $B_c$, the collinear factorization is also a viable approach.
To fully disentangle the contributions from several distinct scales $M_W$,
$m_Q$, $m_Q v$ ($v$ represents the typical velocity of the $c/\bar{b}$
quarks inside $B_c$), and $\Lambda_{\rm QCD}$, we carry out a joint
analysis that combines both light-cone and NRQCD factorization approaches, and compute the
order-$\alpha_s$ correction to $W \to B_c+\gamma$.
With the aid of the celebrated Efremov-Radyushkin-Brodsky-Lepage (ERBL) evolution equation,
we resum the leading collinear logarithms $\alpha_s \ln\! {M_W\over m_Q}$ to all orders in $\alpha_s$.
We also explore some phenomenological implications of our predictions.
%%%%%%%%%%%%%%%%%%%%%%%%%%%%%%%%%%%%%%%%%%%%%%%%%%%%%%%%%%%%%%%%%%%%%%%%%%%%%%
\end{abstract}
%%%%%%%%%%%%%%%%%%%%%%%%%%%%%%%%%%%%%%%%%%%%%%%%%%%%%%%%%%%%%%%%%%%%%%%%%%%%%
\maketitle

%%%%%%%%%%%%%%%%%%%%%%%%%%%%%%%%%%%%%%%%%%%%%%%%%%%%%%%%%%%%%%%%%%%%
\section{Introduction}

Since the discovery of the $B_c$ meson by the \texttt{CDF} Collaboration at Fermilab Tevatron about two decades ago~\cite{Abe:1998wi},
lots of experimental~\cite{Experiments:2012hb} and theoretical~\cite{Chang:1991bp,Chang:1992bb,Chang:1992jb,Chang:1994aw,Chang:1996jt,Chang:2001pm,Wu:2002ig,Chang:2003cr}
efforts have been invested to unravel the properties of this unique
flavored heavy quarkonium, such as its various decay and production rates (for a recent review, see \cite{Brambilla:2010cs}).

The goal of this work is to study the exclusive $B_c$ production in the $W$ decay. 
A huge number of $W$ bosons ($\sim 10^{10}$ events per year) is produced at LHC~\cite{Liao:2011kd}.
Due to the copious yields of $W$, many rare $W$ decay processes may be looked for. 
Among them, numerous hadron production processes in $W$ radiative decay may look
interesting. For instance, the process $W\to D_s\gamma$ has been studied long ago~\cite{Arnellos:1981gy,Keum:1994},
unfortunately with negative result at Tevetron~\cite{Abe:1998vm}.
The $W$ radiative decay into $\pi$ is predicted to have the branching ratio
$10^{-6}-10^{-8}$~\cite{Arnellos:1981gy}, and Tevatron also see null signal~\cite{Albajar:1990qk}.
Recently, a variety of processes about $W$, $Z$ radiative decay into meson have been investigated in 
the light-cone approach~\cite{Grossmann:2015lea}.

Although $W$ radiative decay into $B_c$ is definitely a rare process,  
it may serve as a good playground for enriching our understanding toward QCD.
This process involves several distinct dynamical scales, i.e. the masses
of the $W$ bosons, $b$ and $c$ quark masses, $m_b$ and $m_c$, and the
momentum scale $m_Q v$, and the hadronization scale $\Lambda_{\rm QCD}$.
As a genuine heavy quarkonium, the $B_c$ meson contains several widely separated scales
$M_W \gg m_Q \gg m_{Q} v$. For the hard quarkonium production process, there
exists a well-known perturbative QCD approach, the so-called NRQCD factorization~\cite{Bodwin:1994jh}.
This approach systematically separates the perturbative quantum fluctuations of distance
$1/m_Q$ or shorter, from the nonperturbative effects governing the transition of a heavy quark
pair into a physical quarkonium. 
NRQCD factorization allows one to express the amplitude
for an exclusive quarkonium production process into the sum of product of
short-distance coefficients and the long-distance matrix elements, where the
former can be expanded in powers of strong coupling constant $\alpha_s$,
and the latter long-distance are of nonperturbative origin, whose importance is weighed by
the typical quark velocity $v$.

On the other side, we can also investigate $W\to B_c\gamma$ in the framework
of the light-cone (LC) factorization, also known as the collinear factorization, due to
the mass hierarch $M_W\gg m_Q$. In this approach, the amplitude can be expressed as 
the convolution of the hard-scattering kernel with the nonperturbative 
meson light-cone distribution amplitude (LCDA)

Since this process can be accessed by both NRQCD and LC
factorizations, one may expect an optimized predictions by combining these two approaches
would be desirable. 
A joint analysis would allow the distinct scales $M_W, m_Q, m_Qv$ to be disentangled 
thoroughly. Concretely speaking, we may first express the decay amplitude in the LC factorization
through the next-to-leading order (NLO) in $\alpha_s$.
within the framework of LC factorization. Employing the fact that the $B_c$ LCDA is
not entirely nonperturbative, one can further refactorize it in terms of the perturbatively
calculation function times the NRQCD matrix element.
Through this refactorization program, we can recover the expanded NRQCD prediction simply from
the LC approach. By further employing the evolution equation of the LCDA, we can further resum 
the large collinear logarithms $\alpha_s \ln\tfrac{m_W}{m_Q}$ to all orders in $\alpha_s$.
By this way, we believe that the most precise prediction to the partial width for $W\to B_c\gamma$ 
can be made.

The rest of the paper is organized as follows.
%-----------------------------------
In Sec.~\ref{sec-def}, we set up the notation and decompose the decay amplitude
in terms of two nonvanishing form factors.
%-----------------------------------
In Sec.~\ref{sec-NRQCD},
we calculate the form factors in NRQCD factorization framework at the leading order in velocity,
yet through the NLO in $\alpha_s$. The asymptotic expressions are presented analytically.
%-----------------------------------
Sec.~\ref{sec-LC} is devoted to the calculation of the form factors in the light-cone factorization
through NLO in $\alpha_s$, by implementing the refactorization program for the LCDA of $B_c$.
The connection between LC and NRQCD predictions are elucidated.
Furthermore, we elaborate on how to resum the leading collinear logarithms of type $\ln\frac{M_W}{m_Q}$
to all orders in $\alpha_s$
in the NRQCD short-distance coefficient, with the aid of ERBL evolution equation.
%-----------------------------------
In Sec.~\ref{sec-numerical:results}, we present our numerical results from both NRQCD and light-cone
approaches at NLO in $\alpha_s$. For the latter, we also give the improved prediction with the
resummation effect incorporated.
%-----------------------------------
Finally we summarize in Sec.~\ref{sec-summary}.
%-----------------------------------

%%%%%%%%%%%%%%%%%%%%%%%%%%%%%%%%%%%%%%%%%%%%%%%%%%%%%%%%%%%%%%%%%%%%
\section{Lorentz decomposition of the amplitude \label{sec-def}}

We first set up the kinematics for the process under consideration.
We work in the rest frame of the $W^+$ boson.
We assign the momenta of the outgoing $B_c^+$ and photon by $P$ and $q$, respectively.
We label the polarization vector of $W^+$ by $\varepsilon_W(S_z)$, with $S_z$
the spin-$z$ component, and label the polarization vector of the photon by $\varepsilon_\gamma(\lambda)$,
with $\lambda=\pm 1$ the photon helicity.
The four-momentum of the $W^+$  boson is then $P_W=P+q$.
By the Lorentz invariance, we can decompose the decay amplitude as~\footnote{It is possible to add the third
Lorentz structure $F_3\,\varepsilon_W\cdot q \;
\varepsilon_\gamma^*\cdot P$. Nevertheless, this structure makes vanishing contribution since the photon must be
transversely polarized.}
%-------------------
\begin{equation}
%-------------------
\label{Amp}
%-------------------
\mathcal{A}\PB{W^+ \to B_c^+ + \gamma} = F_1 \; \varepsilon_W\!\PA{\lambda_W}
\cdot \varepsilon_\gamma^*\!\PA{\lambda} + \frac{F_2}{M_W^2}\; i\,\epsilon_{\alpha\beta\gamma\delta}
\varepsilon_W^\alpha\!\PA{S_z} \varepsilon_\gamma^{*\beta}\!\PA{\lambda} P^\gamma q^\delta,
\end{equation}
%-------------------
where $F_1$ and $F_2$ are two scalar form factors, which encode all the nontrivial dynamics.
The main theme of this work is to calculate these two form factors from the
first principles of QCD.

It is convenient to extract the helicity amplitude ${\cal A}_\lambda$ out of \eqref{Amp}.
Since parity invariance is violated by the weak interaction, there are no straightforward
symmetry linking ${\cal A}_{+1}$ with ${\cal A}_{-1}$. Therefore we end up with two
independent helicity amplitudes. One readily infers the decay rates from (\ref{Amp}):
%-------------------
\begin{equation}
%-------------------
\label{Decay:rate:HA}
%-------------------
\Gamma=\frac{1}{2M_W}\frac{|{\bf q}|}{4\pi M_W}{1\over 3}\sum_{\rm Pol} |{\mathcal A}|^2
 =\frac{1}{2 M_W}\frac{|{\bf q}|}{4\pi M_W}\bigg(|{\mathcal A}_+|^2+|{\mathcal A}_-|^2\bigg).
%-------------------
\end{equation}
%-------------------
Equivalently, the unpolarized decay rate can be written as
%-------------------
\beq
%-------------------
\label{Decay:rate:FF}
%-------------------
\Gamma
=\frac{1}{2 M_W}\frac{|{\bf q}|}{4\pi M_W}{1\over 3}s\bigg(2|F_1|^2+2\frac{|{\bf q}|^2}{M_W^2}|F_2|^2 \bigg).
%-------------------
\eeq
%-------------------

%%%%%%%%%%%%%%%%%%%%%%%%%%%%%%%%%%%%%%%%%%%%%%%%%%%%%%%%%%%%%%%%%%%%
\section{The form factors in NRQCD factorization \label{sec-NRQCD}}

In a hard exclusive quarkonium production process, a pair of heavy
quark and an antiquark has to be created in short distance.
In order to have a substantial probability to
form a bound state, the relative motion between
the quark and antiquark have to necessarily be slow. The first condition guarantees that the
asymptotic freedom can be safely invoked to compute the hard-scattering quark-level amplitude
in perturbation theory. The second condition implies that, the quark-level amplitude should
be insensitive to the relative momentum between quark and antiquark, which can thus be expanded
power series of $v$. The binding dynamics is then embedded in those multiplicative
long-distance factors, which are usually modelled by the (derivative of ) quarkonium wave functions
at the origin. These physical picture is at the heart of the NRQCD factorization.

In the NRQCD approach, the amplitude of
$W^+\to B_c^+ \gamma$ at lowest order in $v$ can be expressed as
%-------------------
\begin{equation}\label{Amp_NRQCD0}
\mathcal{A}\!\PB{W\to B_c + \gamma} = \sqrt{\frac{2 M_{B_c} \langle
\mathcal{O}_1 \rangle }{8 N_c m_b m_c}} \mathcal{A}_{Q}\!\PB{W^+\to c\bar{b}(^1S_0^{(1)})+ \gamma},
\end{equation}
%-------------------
where $N_c=3$ is the number of colors, and
$\mathcal{A}_{Q}$ represents the quark-level amplitude with the $B_c$ replaced by the
free $c \bar{b}$ states carrying the same $^1S_0^{(1)}$ quantum number as the leading Fock component of the
$B_c$ meson. $\langle \mathcal{O}_1 \rangle$
denotes the long-distance NRQCD matrix element, which is defined by
%-------------------
\begin{equation}
\langle \mathcal{O}_1 \rangle = \left\vert \langle B_c\vert \psi_c^\dag \chi_b \vert 0
\rangle \right\vert^2 \;,
\end{equation}
%-------------------
where $\psi_c$ is the Pauli spinor field annihilating the $c$, and
$\chi_b^\dagger$ is the Pauli spinor field annihilating the $\bar{b}$ in NRQCD.

For future use, we also introduce the decay constant of $B_c$ via
%-------------------
\begin{equation}\label{decay-constant}
\langle0|{\bar b}\gamma^\mu \gamma^5 c|B_c(p)\rangle=-if_{B_c}\sqrt{2M_{B_c}}p^\mu \;.
\end{equation}
%-------------------

The QCD decay constant in \eqref{decay-constant} can actually be factored into a sum of
the short-distance coefficients
multiplied with the NRQCD matrix elements. At the lowest order in $v$ and $\alpha_s$,
one finds
%-------------------
\begin{equation}
f_{B_c}^{(0)} = \sqrt{\frac{2\langle \mathcal{O}_1 \rangle}{\mbc}} \;.
\end{equation}
%-------------------

To facilitate the future discussions, we strip off some irrelevant
common factors from the form factors, and define the reduced (dimensioneless)
form factors:
%-------------------
\begin{equation}\label{factor-def}
f_{1,2} \equiv  {\sin\theta_W \over e^2 V_{cb} f_{B_c}^{(0)}} \,F_{1,2}\;,
\end{equation}
%-------------------
where $V_{cb}$ is the CKM matrix element, $\theta_W$ signifies the weak mixing angle,
$e$ the electromagnetic coupling, $f_{B_c}^{(0)}$ is the LO decay constant of $B_c$.

%%%%%%%%%%%%%%%%%%%%%%%%%%%%%%%%%%%%%%%%%%%%%%%%%%%
\subsection{NRQCD prediction at LO in $\alpha_s$}

%-------------------------------------------------
\begin{figure}[ht]
\begin{center}
\includegraphics[height=4 cm]{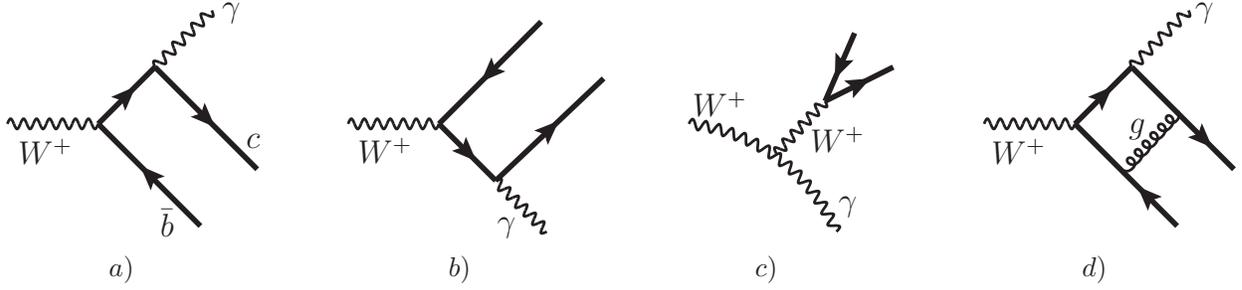}
\caption{\label{Feynman_diag} Some representative diagrams for the quark-level process
$W^+\to c\bar{b}({}^1S_0^{(1)})+\gamma$ through next-to-leading order in $\alpha_s$. }
\end{center}
\end{figure}
%-------------------------------------------------

The procedure of calculating the short-distance coefficients (SDCs) in NRQCD factorization is well known.
The simplest way to proceed is by substituting the physical
$B_c(P)$ with a pair of free heavy quarks.
In Fig.~\ref{Feynman_diag}, we plot three LO diagrams for the quark process
$W^+\to c\bar{b}({}^1S_0^{(1)})+\gamma$.

At LO in $v$, one simply has $M_{B_c} = m_b + m_c$. Furthermore, the momenta of the comoving $c$ and $\bar{b}$ inside the fictitious $B_c$ state are
partitioned with respect to their mass ratio.
Explicitly, $p_c = x_0 P$ and $p_b = \bar{x}_0 P$, with
%-------------------
\beq
%-------------------
x_0 = \frac{m_c}{m_b+m_c},\quad \bar{x}_0\equiv 1-x_0.
%-------------------
\eeq
%-------------------
It is straightforward to prove that both heavy quarks are on their mass shell,
$p_c^2=m_c^2$ and $p_b^2=m_b^2$.

The calculation can be expedited by employing the familiar
spin and color projector technique. Explicitly, one needs project out the
spin- and color-singlet component of the $c\bar{b}$ pair in the quark amplitude:
%-------------------
\begin{equation}\label{proj-spin}
v(p_b) \bar{u}(p_c) \to \frac{1}{\sqrt{8 m_c m_b}} (\gs{p}_b-m_b)
\gamma^5 (\gs{p}_c+m_c) \otimes \frac{\bf{1}}{\sqrt{N_c}},
\end{equation}
%-------------------
where the two spinors are normalized relativistically.

The calculation is quite straightforward, and we just present the LO
results for the reduced form factors~\footnote{In this section, we have focused the process
$W^+\to B_c^+\gamma$. Actually, all the results can be transplanted to
$W^-\to B_c^-\gamma$ case by making the substitution: $e_c\leftrightarrow -e_b$,
$m_c\leftrightarrow m_b$, and $x_0\leftrightarrow \bar{x}_0$.}:
%-------------------
\begin{eqnarray}\label{f-NRQCD-LO}
f_1^{(0)} &=& \frac{1}{4\sqrt{2}} \left( \frac{e_c}{x_0} - \frac{e_b}{\xb_0} \right)
- \frac{1}{2\sqrt{2}}\;, \nn\\
f_2^{(0)} &=& \frac{\mw^2}{2 \sqrt{2}\PA{\mbc^2 - \mw^2}} \left( \frac{e_c}{x_0}
+ \frac{e_b}{\xb_0} \right)\;.
\end{eqnarray}
%-------------------
where the electric charges of charm and bottom quarks are
$e_c = \frac{2}{3}$ and $e_b =- \frac{1}{3}$. Note the last entity in $f_1^{(0)}$
denotes the contribution in Fig.~\ref{Feynman_diag}$c)$, where photon is emitted from
the $W^+$.

In order to make contact with the leading-twist LC prediction, which will be presented in
next section, we are particularly interested in the asymptotic expressions of
(\ref{f-NRQCD-LO}) in the limit $M_W\gg m_Q$:
%-------------------
\begin{eqnarray}\label{NRQCD-T-LC}
f_1^{(0)} &=& \frac{1}{4\sqrt{2}} \left( \frac{e_c}{x_0} - \frac{e_b}{\xb_0} \right)
 - \frac{1}{2\sqrt{2}}, \nn\\
f_2^{(0)} &=& -\frac{1}{2 \sqrt{2}} \left( \frac{e_c}{x_0} + \frac{e_b}{\xb_0} \right).
\end{eqnarray}
%-------------------
It is reassuring to note that with the identification $m_b\to m_c$ and $e_b\to e_c$ made,
one correctly reproduces the corresponding pattern for $\gamma^*\to \eta_c\gamma$:
$f_1{(0)}$ is vanishing and only $f_2^{(0)}$ survives.

In literature many authors may regard the $B_c$ as the heavy-light meson, to which the celebrated
heavy quark effective theory (HQET) may apply.
We do not hold this viewpoint. On the physical ground, we believe $B_c$ is a genuine quarkonium rather
than a heavy-light meson.
Nevertheless, it may be theoretically enlightening to look at the asymptotic expressions of $f_{1,2}$ in
\eqref{f-NRQCD-LO} in the HQET limit, {\it i.e.},  $M_W\sim m_b\gg m_c$:
%-------------------
\begin{eqnarray}
f_1^{(0)} &=& \frac{1}{4\sqrt{2}} \frac{e_c}{x_0}, \nn\\
f_2^{(0)} &=& \frac{r^2}{2 \sqrt{2}\PA{1 - r^2}} \frac{e_c}{x_0},
\end{eqnarray}
%-------------------
where
%-------------------
\begin{equation}
r = \frac{\mw}{m_b} \;,\; x_0 \simeq \frac{m_c}{m_b}\;.
\end{equation}
%-------------------
Only Fig.~\ref{Feynman_diag}$a)$ with photon emitted from the $c$ line survives in this limit.
We recover the same power behavior $1/x_0$ as what arises from the heavy-quark recombination
mechanism~\cite{Braaten:2001bf,Braaten:2002yt}.

In our case, $m_b$ is not much greater than $m_c$, so there is no need
to separate these two scales in the present work. Nevertheless, for the hard exclusive process
involving heavy-light hadron such as the $B$ and $D$ mesons,
it is physically meaningful to factorize these two scales within the HQET.

%%%%%%%%%%%%%%%%%%%%%%%%%%%%%%%%%%%%%%%%%%%%%%%%%%%
\subsection{NRQCD prediction at NLO in $\alpha_s$}

We then proceed to compute NLO perturbative corrections to the reduced
form factors $f_{1,2}$. There are totally ten NLO Feynman graphs for
$W^+\to c\bar{b}({}^1S_0^{(1)})+\gamma$ , one of which is
depicted in Fig.~\ref{Feynman_diag}.
We conduct the calculation in Feynman gauge, and employ dimensional regularization
to regularize both UV and IR divergences.
After renormalization procedure, we end up with the UV and IR finite NRQCD SDCS $f_{1,2}^{(1)}$.

Since $f_{1,2}^{(1)}$ involves three distinct mass scales: $M_W$, $m_b$ and $m_c$, it is conceivable that
their analytical expressions are rather cumbersome, which are considerably more lengthy
than $\gamma^*\to \eta_c\gamma$. Therefore we refrain from providing their complete expressions here.
However, for the sake of clarity, we decide to present the asymptotic expressions of $f_{1,2}^{(1)}$
in two kinds of distinct of limits:
\begin{itemize}
\item Expansion at LC limit ($M_W\gg m_Q$):
\begin{eqnarray}\label{NLO NRQCD}
f_1^{(1)} &=& \Bigg\{ - \frac{C_F}{16 \sqrt{2}} \frac{e_c}{x_0} \Bigg[
(2 \ln x_0+3) \left( \ln\!\frac{\mbc^2}{\mw^2} + i \pi\right)\nonumber\\
&&+\ln^2x_0+\left(\frac{3\,x_0}{\bar{x}_0}+5\right) \ln x_0+3\ln\bar{x}_0-4{\rm Li}_2(x_0)+\frac{2\pi^2}{3} + 9 \Bigg] \nn\\
&& - \PA{\exchange} \Bigg\} -\frac{3 C_F }{8 \sqrt{2}} \left[ (x_0-\bar{x}_0) \ln\frac{x_0}{\bar{x}_0}-2\right] + \cdots \;, \nn\\
f_2^{(1)} &=& \frac{C_F}{8 \sqrt{2}} \frac{e_c}{x_0} \Bigg[
(2 \ln x_0+3) \left( \ln\!\frac{\mbc^2}{\mw^2} + i \pi\right)+\ln^2x_0+\left(\frac{x_0}{\bar{x}_0}+5\right) \ln x_0 \nn\\
&&+3\ln\bar{x}_0-4{\rm Li}_2(x_0)+\frac{2\pi^2}{3} + 9 \Bigg] + \PA{\exchange} + \cdots \;,
\end{eqnarray}
\item Expansion at HQET limit ($M_W\sim m_b\gg m_c$):
\begin{eqnarray}
f_1^{(1)} &=& -\frac{C_F}{16 \sqrt{2} r^2} \frac{e_c }{x_0} \Bigg[ r^2
\ln^2x_0 + r^2 \PB{-2\ln(r^2 -1)+2 i \pi +5} \ln x_0 \nonumber\\
&& + (-3 r^2 +1) \ln(r^2 -1) + 2 r^2  \text{Li}_2\left(\frac{r^2 -1}{r^2 }\right) + r^2  \ln^2\left(\frac{r^2}{r^2-1}\right) \nonumber\\
&& + \PA{9+\frac{\pi^2}{3}+3i\pi} r^2+2 i \pi r^2 \ln\left(\frac{r^2 }{r^2 -1}\right) - i \pi  \Bigg],
\nn\\
f_2^{(1)} &=& \frac{2r^2}{1-r^2} f_1^{(1)}.
\end{eqnarray}
\end{itemize}

In this work, we will be exclusively interesting with the LC limit \eqref{NLO NRQCD}.
The asymptotic expression of $f_2^{(1)}$ in such limit, with the substitutions
$e_b\to e_c$ and $m_b\to m_c$ made, coincides with the
asymptotic expression (33) found for $\gamma^*\to \eta_c+\gamma$~\cite{Sang:2009jc}.
In the next section, we will demonstrate
(\ref{NLO NRQCD}) can be exactly reproduced by the NLO calculation in the
light-cone factorization by implementing the refactorization program.

%%%%%%%%%%%%%%%%%%%%%%%%%%%%%%%%%%%%%%%%%%%%%%%%%%%
\section{Form factors in light-cone approach with the
refactorization program \label{sec-LC}}

At asymptotically large $M_W$, the outgoing $B_c$ moves nearly with the speed of
the light. By virtue of the asymptotic freedom,
the hard-scattering quark amplitude can be accessed by perturbation theory.
Since both the quark and antiquark inside $B_c$ are dictated by the light-like
kinematics, the hard-scattering amplitude is insensitive to the
the quark mass, $m_{Q}$, as well as the transverse momentum carried by the quark and antiquark, $p_\perp$. 
Thus, the amplitude can be expanded in powers of $p_\perp$ and $m_{Q}$,
while the nonperturbative wave function together with the $p_\perp$ and quark-mass-dependent
effects can be lumped into the LCDAs of $B_c$. This picture naturally results in the
collinear factorization.

At the leading power in $1/M_W$, the reduced form factors $f_{1,2}$ can be expressed as
the convolution of the perturbatively calculable hard-scattering amplitude $T_{1,2}$ with the
leading-twist LCDAs of $B_c$, signified by $\Phi_{B_c}(x)$:
%-------------------
\begin{equation}\label{leading:twist:factorization:theorem}
f_{1,2}^{\rm LC}=\frac{1}{f_{B_c}^{(0)}}\int_0^1 \!\! dx \, T_{1,2}(x,Q^2,\mu^2_R,\mu^2_F)\,
\Phi_{B_c}^*(x,\mu^2_F)+ {\mathcal O}(1/Q^2)\,,
\end{equation}
%-------------------
where $x$ represents the fractions of light-cone momentum
carried by the $c$ quark in the $B_c$ state,
and $\mu_R$, $\mu_F$ denote the renormalization and factorization
scales, respectively. 

The $B_c$ LCDA $\Phi_{B_c}(x)$ is defined through
%-------------------
\begin{equation}\label{Phi-def}
\Phi_{B_c}(x)\equiv \int\!\!\frac{dz^-}{2\pi} e^{i x p^+z_-}
\langle 0|\bar{b}(z_-)[z_-,0]\gamma^+\gamma^5 c(0)|B_c(p)\rangle
\equiv-if_{B_c}\hat{\phi}(x),
\end{equation}
%-------------------
where $[z_-,0]$ denotes the gauge link
%-------------------
\begin{equation}
[z_-,0]={\mathcal P}{\rm exp}\bigg[ig_s\int_0^{z_-}\!\! dx A^+(x)\bigg].
\end{equation}
%-------------------

Combining (\ref{decay-constant}) and (\ref{Phi-def}), it is ready to obtain the
normalization relation: $\int_0^1 dx\phi(x)=1$.
Since we have extracted the LO decay constant $f_{B_c}^{(0)}$ in (\ref{factor-def}),
the prefactor $\tfrac{1}{f_{B_c}^{(0)}}$ should be put in (\ref{leading:twist:factorization:theorem}).

The LCDA of a quarkonium is quite different with that of light meson.
The latter is sensitive to the effect at the scale
$\Lambda_{QCD}$, so is a genuinely nonperturbative quantity, which
should be tackled by, i.e. lattice QCD, or phenomenological methods.
By contrast, in the spirit of refactorization~\cite{Ma:2006hc,Bell:2008er},
the perturbative LCDA $\hat{\phi}$ for a quarkonium
may be viewed as the short-distance coefficient function associated with
matching the quarkonium LCDA onto the NRQCD vaccum-to-quarkonium matrix element,
while the nonperturbative effect of order $m_Q v$ or lower
has been entirely encoded in $f_{B_c}$.
Since heavy quark mass can serve as an infrared cutoff,
this coefficient function, which captures the effects of collinear modes with virtuality of order
$m_{Q}^2\gg \Lambda_{\rm QCD}^2$,
can be reliably accessible by perturbation theory owing to asymptotic freedom.
At the LO in $v$, this perturbative function can be expanded in power series of $\alpha_s$:
%-------------------
\begin{eqnarray}
\hat{\phi}(x,\mu_F^2) &=& \hat{\phi}^{(0)}(x)+ {\alpha_s(\mu_F^2)
\over \pi} \hat{\phi}^{(1)}(x,\mu_F^2)+ \cdots.
\label{Bc:jet:function:expansion}
\end{eqnarray}
%-------------------
The LO coefficient function can be trivially inferred,
%-------------------
\beq \label{LO:pert:LCDA}
\hat{\phi}^{(0)}(x)=\delta(x-x_0),
\eeq
%-------------------
which simply reflects that both $c$ and $\bar{b}$ partition the
momentum of $B_c$ commensurate to their mass ratios. This is
compatible with the LO NRQCD expansion, in that $c$ and $\bar{b}$
are at rest relative to each other in the $B_c$ rest frame~\footnote{Note that the 
LCDA of $B_c$ is asymmetric about
$x={1\over 2}$, which is rather different from the $\pi$ case. 
That is because the flavor symmetry
between heavy quarks $c$ and $b$,  unlike the isospin symmetry between $u$ and $d$,
is badly broken.}.

On the other side, the decay constant $f_{B_c}$ can also be expanded in power series of
$\alpha_s$:
%-------------------
\begin{eqnarray}\label{decay-constant-expand}
f_{B_c} &=& f^{(0)}_{B_c} \left( 1+ {\alpha_s(M^2_{B_c})\over \pi} \mathfrak{f}_{B_c}^{(1)}
+\cdots \right),
\end{eqnarray}
%-------------------
where we neglect the contributions at higher order in $v$.
According to (\ref{decay-constant}), the hard matching coefficient of
$f_{B_c}$ can be obtained through matching the QCD axial-vector current to
its NRQCD counterpart order by order in $\alpha_s$. In this work,
$\mathfrak{f}_{B_c}^{(1)}$ is sufficient for our purpose and we will present the
explicit result in the following subsection.

Analogously, we can split the reduced form factors in the LC factorization as
%-------------------
\begin{eqnarray}
f^{\rm LC}_{1,2}=f^{{\rm LC}(0)}_{1,2}+\frac{\alpha_s}{\pi}f^{{\rm LC}(1)}_{1,2}+\cdots.
\end{eqnarray}
%-------------------

By using the notations introduced in the preceding paragraphs, we
are able to reexpress the factorization formula (\ref{leading:twist:factorization:theorem})
as
%-------------------
\begin{eqnarray}\label{LC-factorization}
f^{{\rm LC}(0)}_{1,2}&\sim& T^{(0)}\otimes\hat{\phi}^{(0)},\nn\\
f^{{\rm LC}(1)}_{1,2}&\sim& T^{(1)}\otimes\hat{\phi}^{(0)}+T^{(0)}\otimes\hat{\phi}^{(1)}
+\mathfrak{f}_{B_c}^{(1)}T^{(0)}\otimes\hat{\phi}^{(0)},
\end{eqnarray}
%-------------------
where $\otimes$ indicates the convolution, and a common factor has been suppressed.

In the following, we will explicitly carry out the light-cone calculation outline
in (\ref{LC-factorization}). The whole light-cone program with refactorization implemented here,  
is very similar to that for $B_c$ electromagnetic form factor~\cite{Jia:2010fw}. 

%%%%%%%%%%%%%%%%%%%%%%%%%%%%%%%%%%%%%%%%%%%%%%%%%%%
\subsection{LC prediction in LO in $\alpha_s$}

We first compute the hard-scattering kernel at LO. 
The calculation is again facilitated with the covariant projector, 
except we set the mass and momenta of the heavy quarks in (\ref{proj-spin}) as
%-------------------
\begin{eqnarray}
m_Q=0, p_c=x P, p_b=\bar{x} P,
\end{eqnarray}
%-------------------
where $\bar{x}\equiv 1-x$.
Consequently, (\ref{proj-spin}) reduces to
%-------------------
\begin{eqnarray}\label{proj-spin-LC}
\Pi_0=\sqrt{\frac{x_0\bar{x}_0}{2}}
\gamma^5 \gs{p} \otimes \frac{\bf{1}}{\sqrt{N_c}}. \;
\end{eqnarray}
%-------------------

It is a straightforward exercise to obtain the Born order hard-scattering kernel:
%-------------------
\begin{eqnarray}\label{LC-T-LO}
T_1^{(0)}(x, \mu_R^2, \mu_F^2) &=& \frac{1}{4\sqrt{2}} \PA{ \frac{e_c}{x} -
 \frac{e_b}{\xb} } - \frac{1}{2\sqrt{2}}, \nn\\
T_2^{(0)}(x, \mu_R^2, \mu_F^2) &=& -\frac{1}{2\sqrt{2}} \PA{ \frac{e_c}{x} +
\frac{e_b}{\xb} }.
\end{eqnarray}
%-------------------
With these results and $\hat{\phi}^{(0)}$ in \eqref{LO:pert:LCDA}, 
we immediately predict
the reduced form factors $f_{1,2}^{{\rm LC},(0)}$:
%-------------------
\begin{eqnarray}
f_1^{{\rm LC},(0)} &=& \frac{1}{4\sqrt{2}} \left( \frac{e_c}{x_0} -
\frac{e_b}{\xb_0} \right) - \frac{1}{2\sqrt{2}}, \nn\\
f_2^{{\rm LC},(0)} &=& -\frac{1}{2 \sqrt{2}} \left( \frac{e_c}{x_0} + \frac{e_b}{\xb_0} \right).
\end{eqnarray}
%-------------------
Not surprisingly, $f_{1,2}^{{\rm LC},(0)}$ exactly recovers Eq.~(\ref{NRQCD-T-LC}), which
is obtained from the asymptotic expression from NRQCD approach.

%%%%%%%%%%%%%%%%%%%%%%%%%%%%%%%%%%%%%%%%%%%%%%%%%%%
\subsection{LC prediction at NLO in $\alpha_s$}

We continue to investigate the NLO perturbative correction.
To this end, an important piece of knowledge in (\ref{LC-factorization}) 
is the NLO perturbative correction to the hard-scattering kernel.
Fortunately, $T_{1,2}^{(1)}$ can be largely transplanted from a preceding NLO calculation
for $\gamma^*\gamma\to \pi$ transition form factor~\cite{Braaten:1982yp}.
The $\gamma_5$ prescription in dimensional regularization causes some subtlety, where
the author of \cite{Braaten:1982yp} finds some finite counter-term must be imposed
to satisfy the axial Ward identity.
In the so-called naive dimensional regularization (NDR) scheme, one has
%-------------------
\begin{eqnarray}
T_1^{(1)}(x, \mu_F^2) &=& \frac{C_F}{16 \sqrt{2}} \frac{e_c}{x}
\Bigg[ -(2 \ln x+3) \ln\!\frac{-\mu_F^2}{\mw^2} + \ln^2x
- \frac{3\,x}{\bar{x}}
\ln x - 9 \Bigg] - \PA{\xchange},\nn\\
T_2^{(1)}(x, \mu_F^2) &=& -\frac{C_F}{8 \sqrt{2}} \frac{e_c}{x}
\Bigg[  -(2 \ln x+3) \ln\!\frac{-\mu_F^2}{\mw^2} + \ln^2x
- \frac{x}{\bar{x}} \ln x - 9 \Bigg]
+ \PA{\xchange},
\end{eqnarray}
%-------------------
whereas in the so-called 't Hooft-Veltman (HV) scheme, one has
%-------------------
\begin{eqnarray}
T_1^{(1)}(x, \mu_F^2) &=& \frac{C_F}{16 \sqrt{2}} \frac{e_c}{x}
\Bigg[ -(2 \ln x+3) \ln\!\frac{-\mu_F^2}{\mw^2} + \ln^2x
+ \frac{5\,x}{\bar{x}}
\ln x - 9 \Bigg] - \PA{\xchange}+\frac{C_F}{2\sqrt{2}} \;,\nn\\
T_2^{(1)}(x, \mu_F^2) &=& -\frac{C_F}{8 \sqrt{2}} \frac{e_c}{x}
\Bigg[ -(2 \ln x+3) \ln\!\frac{-\mu_F^2}{\mw^2} + \ln^2x
+ \frac{7x}{\bar{x}} \ln x - 9 \Bigg]
+ \PA{\xchange} \;.
\end{eqnarray}
%-------------------

In the HV scheme, there is no unambiguity associated with the trace operation.
Since there is no axial current entering the amplitude, we need not pay extra attention to the counterterm just as
for the process $\gamma^*\to \eta_c+\gamma$. 
In NDR scheme, we just move the $\gamma_5$ originating from the projection operator to the middle (sandwiched 
by the two gluons vertices) in the box diagrams, and keep $\gamma_5$ in the outside in other diagrams(otherwise an extra
finite counterterm is mandatory). This is what we call NDR scheme for this case. 
In our opinion, the manipulation adopted in the hard-scattering kernel must be consistent 
with the corresponding operation in computing the LCDA, 
where $\gamma_5$ stays in the middle of the box diagrams~\footnote{For a more rigorous treatment
of $\gamma_5$ ambiguity in light-cone approach, using the evanescent operators rather than the
projector approach, we refer the interested readers to \cite{Wang:2013ywc,Wang:2018wfj}.}.

According to (\ref{LC-factorization}), another important class of NLO correction stems from
convoluting the tree-level hard-scattering kernel with the NLO perturbative LCDA of $B_c$.
A nice feature in our light-cone treatment is that, this perturbative LCDA  
can be systematically accessed in perturbation theory.
It is the very feature that makes the refactorization program
practically useful.

Fortunately, the NLO perturbative LCDA for the $B_c$ meson has already been calculated by 
Bell and Feldmann~\cite{Bell:2008er}:
%---------------------------------
\bqa\label{LCDA-NLO}
%---------------------------------
& & \hat{\phi}^{(1)}(x,\mu_F^2)=
{C_F\over 2} \bigg\{\bigg(\ln{\mu^2_F\over M_{B_c}^2 (x_0-x)^2}-1 \bigg)
\bigg[
{x_0+\bar{x}\over x_0-x} {x\over x_0}
\theta(x_0- x) +\Delta\frac{4x}{x_0}\theta(x_0-x)\nn\\
&+& \bigg(
%---------------------------------
\begin{array}{c}
%---------------------------------
x\leftrightarrow \bar{x}
%---------------------------------
\\
%---------------------------------
x_0 \leftrightarrow \bar x_0
%---------------------------------
\end{array}
%---------------------------------
\bigg)\bigg]\bigg\}_+
%---------------------------------
%---------------------------------
+ C_F \Bigg\{
\bigg({x \bar{x}\over (x_0-
x)^2}\bigg)_{++} +  {1\over 2}\,\delta^\prime(x-x_0)
\bigg(2 x_0 \bar{x}_0 \ln {x_0 \over \bar{x}_0}+
x_0-\bar{x}_0\bigg)\Bigg\}\,,\phantom{xx}
%---------------------------------
\label{Bc:jet:function:NLO}
%---------------------------------
\eqa
%---------------------------------
and
%-------------------
\begin{eqnarray}
f^{(1)}_{B_c}=f^{(0)}_{B_c}\bigg[1+\frac{\alpha_s C_F}{4\pi}\bigg(-6+3(x_0-\bar{x}_0)\ln\frac{x_0}{\bar{x}_0}\bigg)+4\Delta\bigg],
\end{eqnarray}
%-------------------
where $\Delta=0$ for the NDR scheme, and $\Delta=1$ for the HV scheme~\cite{Wang:2013ywc}.

In (\ref{LCDA-NLO}), the ``+'' and ``++"-prescriptions are understood in the sense
of distributions. For a test function $f(x)$ which is smooth near $x=x_0$, 
its convolutions with the ``+'' and ``++"-functions are given by~\cite{Bell:2008er}
%---------------------------------
\begin{subequations}
%---------------------------------
\bqa
%---------------------------------
&&\int_0^1 \!\! dx\, [g(x)]_+  f(x)\equiv \int_0^1\!\! dx\,
g(x) \left(f(x)-f(x_0)\right)\,,
%---------------------------------
\\
%---------------------------------
&&\int_0^1 \!\! dx\, [g(x)]_{++} f(x) \equiv \int_0^1\!\! dx\,
g(x) \bigg( f(x)-f(x_0)-f^\prime(x_0)(x-x_0)\bigg)\,.
%---------------------------------
\eqa
%---------------------------------
\end{subequations}
%---------------------------------

The last missing piece is the NLO perturbative correction to the decay constant $f_{B_c}$, namely
$\mathfrak{f}_{B_c}^{(1)}$, which was computed by Braaten and Fleming in 1995~\cite{Braaten:1995ej}:
%---------------------------------
\bqa
%---------------------------------
\mathfrak{f}_{B_c}^{(1)} &=& -{3\over 2}C_F + {3\over 4} C_F
(x_0-\bar{x}_0) \ln {x_0\over\bar{x}_0},
%---------------------------------
\label{Bc:decay:const:matching:coef:NLO}
%---------------------------------
\eqa
%---------------------------------
which is symmetric under $x_0\leftrightarrow \bar{x}_0$.

It is the time to piece all the relevant elements together.
It is a straightforward exercise to verify that,
by adding three terms in (\ref{LC-factorization}) together, we do
reproduce the expanded NRQCD prediction (\ref{NLO NRQCD}) successfully.

\subsection{Resummation of leading logarithms using evolution equation}

One of the great virtues of the light-cone program is that we can utilize the renormalization group
equation (RGE) to efficiently resum the large collinear logarithms of type $\alpha_s\ln M_W^2/M_{B_c}^2$. 
This is impossible in NRQCD approach. This type of resummation for hard quarkonium production was
first illustrated in \cite{Jia:2008ep}. 
Recently there have been some development for the numerical algorithm about resummation~\cite{Bodwin:2016edd,Bodwin:2017pzj}.
  
As is well known, the evolution of the leading-twist LCDA of a
meson is governed by the celebrated Efremov-Radyushkin-Brodsky-Lepage
(ERBL) evolution equation~\cite{Efremov:1979qk,Lepage:1979zb}.
Particularly, the perturbative LCDA for a $B_c$ meson
obeys the following evolution equation:
%------------------
\bqa
%------------------
{d\over d \ln \mu_F^2} \hat{\phi}(x,\mu_F^2)
 &=&
{\alpha_s(\mu_F^2)\over \pi} \, \int^1_0 \! dy \, V_0(x,y)\,\hat{\phi}(y,\mu_F^2)+ O(\alpha_s^2),
%------------------
\label{BL-evolution:eqn:spin:zero}%
%------------------
\eqa
%------------------
where
%------------------
\bqa
%------------------
V_0(x,y) &=& {C_F\over
2} \left[{1-x\over 1-y}\left(1+{1\over
x-y}\right)\theta(x-y)+ {x\over y}\left(1+{1\over
y-x}\right)\theta(y-x)\right]_+
%------------------
\label{BL-kernel-pseudoscalar}%
%------------------
\eqa
%------------------
is the corresponding evolution kernel. 
Upon substituting (\ref{Bc:jet:function:NLO}) into
(\ref{Bc:jet:function:expansion}), one can explicitly verify that
(\ref{BL-evolution:eqn:spin:zero}) is indeed satisfied.

As is well known, the kernel $V_0$ admits the eigenfunctions $G_n(x)
\equiv x(1-x) C_n^{(3/2)}(2x-1)$, which are Gegenbauer polynomials
of rank ${3\over 2}$ multiplied by the weight function $x(1-x)$:
%------------------
\begin{subequations}
%------------------
\bqa
%------------------
\int^1_0 \!dy \, V_{0}(x,y) \, G_n(y) &=& -\gamma_n \, G_n(x),
%------------------
\label{V0:eigenfunction}
\\
%------------------
%G_n(x) & \equiv & x(1-x) C_n^{(3/2)}(2x-1),
%%------------------
%\label{V0:eigenfunction}
%\\
%------------------
\gamma_n &=& {1\over 2} + 2 \sum_{j=2}^{n+1} {1\over j}- {1
 \over (n+1)(n+2)}\,.
%------------------
\label{Gegenbauer:eigenvalue}
%------------------
\eqa
%------------------
\end{subequations}
%------------------
It is a standard practice to expand the $\hat{\phi}(x)$ in the
basis of $G_n$:
%------------------
\begin{subequations}
\label{gegenbauer-expansion}
%------------------
\bqa
%------------------
\hat{\phi}(x,\mu) &=& \sum_{n=0}^\infty \hat{\phi}_n(\mu) \, G_n(x),
%------------------
\\
%------------------
\hat{\phi}_n(\mu) &=&  {4(2n+3)\over (n+1)(n+2)}\int_0^1 \! dx \,
C_n^{(3/2)}(2x-1)\,\hat{\phi}(x,\mu).
%------------------
%------------------
\eqa
%------------------
\end{subequations}
%------------------

Substituting (\ref{gegenbauer-expansion}) into (\ref{V0:eigenfunction}),
we immediately get the solutions of RG equation for each Gegenbauer moment:
%------------------
$\hat{\phi}_n(\mu)$
\begin{equation}
\hat{\phi}_n(Q^2)= \hat{\phi}_n(m^2) \left[ \frac{\alpha_s(Q^2)}{\alpha_s(m^2)}
 \right]^{d_n},
\end{equation}
where
%------------------
\begin{equation}
%------------------
d_n = \frac{2C_F \gamma_n}{\beta_0} = \frac{2C_F}{\beta_0} \left[ \frac{1}{2} + 2
\sum_{j=2}^{n+1}\frac{1}{j} - \frac{1}{(n+1)(n+2)} \right].
%------------------
\end{equation}
%------------------

At leading logarithmic (LL) accuracy, the reduced form factors can be expressed as 
the sum of Gegenbauer moments:
\begin{eqnarray}\label{LL-resum-1}
f_{1,2}^{LL} &=&\int_0^1 dx \; T_{1,2}^{(0)}(x) \, \hat{\phi}^{(0)} (x, Q^2)\nn\\
&=& \sum_{n=0}^{\infty} \frac{4(2n+3)}{(n+1)(n+2)} C_n^{(3/2)}(2x_0-1)
\left[ \frac{\alpha_s(Q^2)}{\alpha_s(m^2)} \right]^{d_n} \int_0^1 dx \;
T_{1,2}^{(0)}(x) \, G_n(x),
\end{eqnarray}
where we have taken $\hat{\phi}^{(0)}(x, m^2) = \delta(x-x_0)$.

In our case, $T_{1,2}^{(0)}(x)$ in \eqref{LC-T-LO} can be generically parameterized as
%------------------
\begin{equation}
%------------------
T_{1,2}^{(0)}(x) = c_0 + \frac{c_1}{x} + \frac{c_2}{\bar{x}} \,,
\end{equation}
%------------------
where $c_0$, $c_1$ and $c_2$ are $x$-independent coefficients. With the aid of the
following identities:
%------------------
\begin{eqnarray}
%------------------
\int_0^1 dx \; G_{n}(x) &=& \left\{
\begin{array}{cl}
\displaystyle\frac{1}{6} & (n=0) \\
\\
0 & (n>0)
\end{array}
\right. ,\nn\\
\int_0^1 dx \; \frac{G_{n}(x)}{x} &=& \frac{1}{2} (-1)^n, \nn\\
\int_0^1 dx \; \frac{G_{n}(x)}{\bar{x}} &=& \frac{1}{2},
%------------------
\end{eqnarray}
%------------------
we can reach the final expressions for the form factors:
%------------------
\begin{equation}
%------------------
\label{LL-resum-2}
%------------------
f_{1,2}^{\rm LL} = \PC{ c_0 + \sum_{n=0}^{\infty} \frac{4(2n+3)}{(n+1)(n+2)}
C_n^{(3/2)}(2x_0-1) \left[ \frac{\alpha_s(Q^2)}{\alpha_s(m^2)} \right]^{d_n}
\left[ (-1)^n \frac{c_1}{2} + \frac{c_2}{2} \right] }.
%------------------
\end{equation}
%------------------
This formula can be directly employed to resum the leading logarithms numerically.

\subsection{The optimized LC prediction}

It is possible to make an optimized prediction by combining the light-cone and NRQCD approach.
We can start from the fixed NLO prediction in \eqref{NLO NRQCD}, then combine with \eqref{LL-resum-2}
to take into account the leading-logarithm resummation. Of course, we should subtract the 
$\alpha_s \ln\!\frac{\mbc^2}{\mw^2}$ terms in $f_{1,2}^{1}$ to avoid the double counting.
Provided that $M_W\gg M_{B_c}$, this procedure generates the most precise and optimal prediction.

%%%%%%%%%%%%%%%%%%%%%%%%%%%%%%%%%%%%%%%%%%%%%%%%%%%
\section{Numerical results \label{sec-numerical:results}}

In current section,  we first demonstrate the merit of light-cone 
refactorization compared with the fixed-order NRQCD prediction.

\begin{figure}[ht]
\begin{center}
\includegraphics[height=5cm]{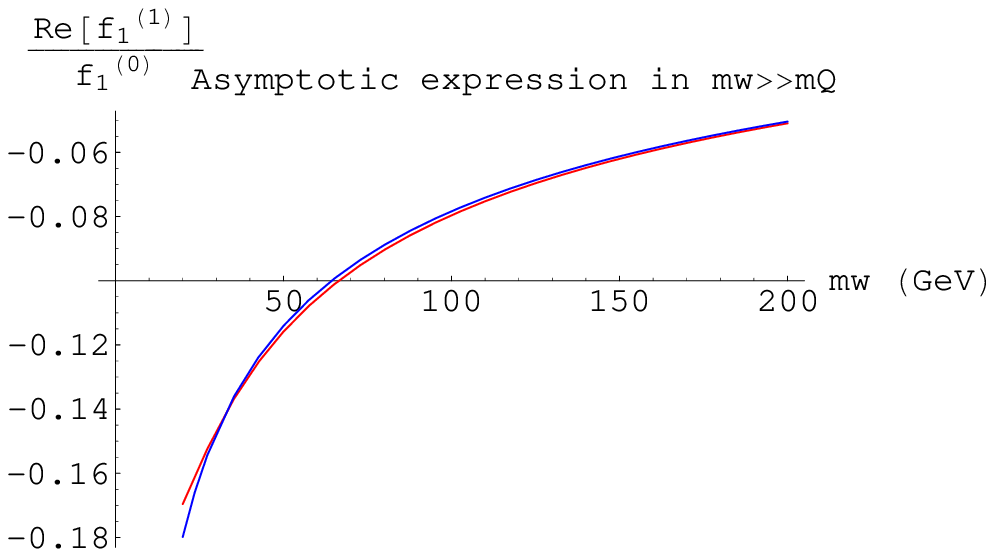}
\includegraphics[height=5cm]{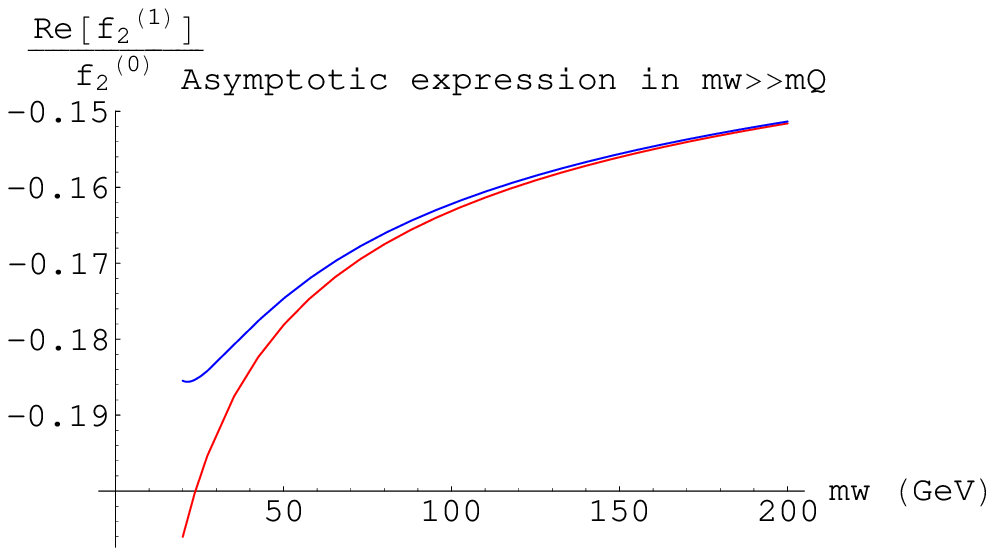}
\caption{\label{fig-lc-1} The NRQCD and LC predictions for the reduced form factors 
as the function of $M_W$. 
The red line represents
the NRQCD one, while blue line represents another. In the figure, we have divided
the NLO results by the LO NRQCD predictions. In addition, we take $m_b=4.6$ GeV
and $m_c=1.4$ GeV.}
\end{center}
\end{figure}

First, to show at which circumstance the LC limit ($M_W\gg m_Q$) can be reliably used,
we plot the curves of the full NRQCD and asymptotic expressions as the
function of $M_W$ for the two form factors $f_{1, 2}$ in Fig.~\ref{fig-lc-1}.
%-------------------------------------------------
%-------------------------------------------------
From the figure, we see that the LC asymptotic expressions for both form factors
converge to the NRQCD ones even at relatively low $M_W$, i.e. $M_W=30$ GeV. 
This is particularly true for $f_1$.

\begin{figure}[ht]
\begin{center}
\includegraphics[height=5cm]{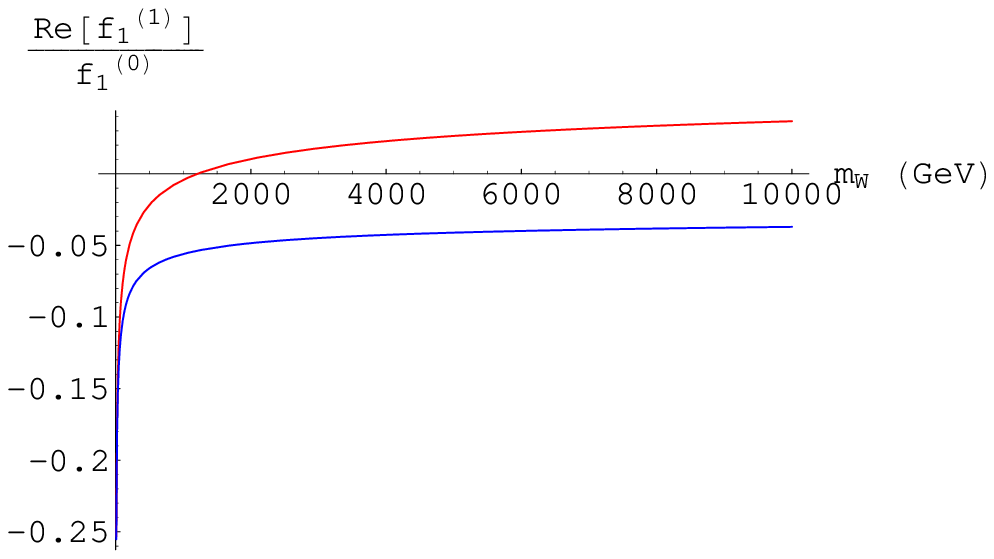}
\includegraphics[height=5cm]{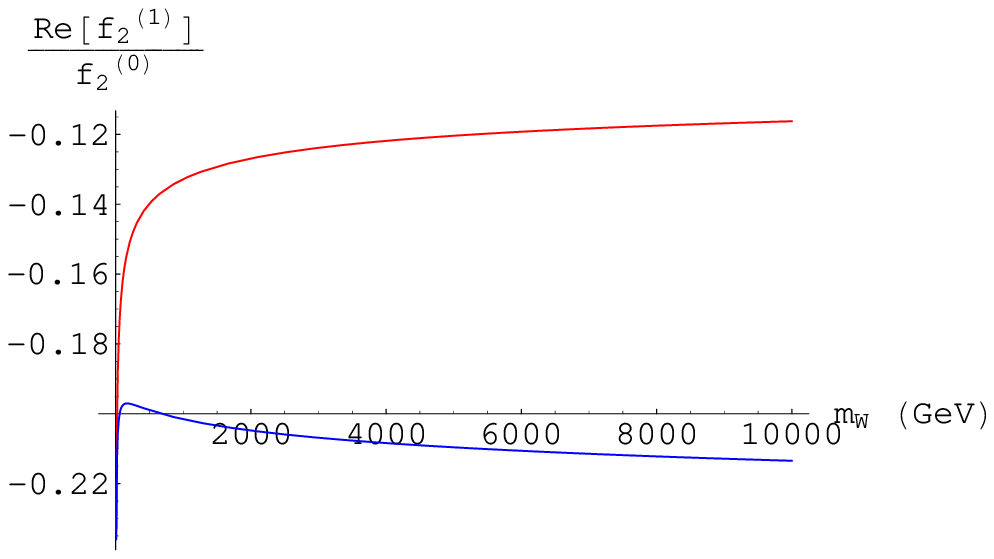}
\caption{\label{fig-lcresum-1} The NLO LC predictions without and with the resummation of leading collinear logarithms. 
The red line represents the one without resummation, while the blue line including the resummation. 
We have divided the NLO results by the LO LC predictions. We again take $m_b=4.6$ GeV
and $m_c=1.4$ GeV.}
\end{center}
\end{figure}

Next, we show how the LL resummation improves the theoretical
prediction in Fig.~\ref{fig-lcresum-1}.
%-------------------------------------------------
%-------------------------------------------------
As can be seen in the figrue, at the region of relative
large $M_W$, i.e. $M_W>2$ TeV, the resummed results will improve the LC prediction smoothly,
due to the fact that the strong coupling constant is asymptotically free. Compared with
$f_2$, resummation for $f_1$ is relatively important in magnitude. This is because
there exists a cancelation in the `constant' term in the expression of $f_2$.

Finally, we present the predictions for the cross section of the process $W^+\to B_c^+\gamma$.
We take the heavy quark masses to be $m_c=1.4$ GeV and $m_b=4.6$ GeV, and the mass
of the $W$ boson to be $M_W=80.6$ GeV. The NRQCD matrix elements are chosen to
be $\langle{\mathcal O}_1\rangle=2N_c \psi(0)^2=0.785\; {\rm GeV}^3$,
translated from the wave function at the origin in Cornell potential model~\cite{Eichten:1995ch}.
In the NRQCD approach, we then obtain $\Gamma^{(0)}=1.77$ eV, and the ${\cal O}(\alpha_s)$ correction becomes negative$\Gamma^{(1)}=-3.55\alpha_s$ eV.
By taking $\alpha_s=0.12$, we find the NLO perturbative corrections to be around $-23\%$, and the
partial width through the NLO accuracy to be $1.34$ eV. 

In the light-cone approach, after incorporating the LC resummation, we obtain the most optimal
prediction to be $1.48$ eV.
We believe it is by far the most precise model-independent prediction for this decay process.

The full width of $W$ boson is $\Gamma_W = 2.085\pm 0.042$ GeV~\cite{Tanabashi:2018oca}.
The branching fraction of this $W$ radiative decay channel is less than $10^{-9}$, corresponding to an extremely rare
process. Although a tremendous amount of $W$ events have already been accumulated at
the \texttt{LHC} experiments, concerning the difficulty of reconstructing the $B_c$ signal and the detector
efficiency, it still appears to be an impossible mission to observe this rare decay signal
in the foreseeable future. We hope the potential next-generation high-energy collider such
as \textsf{SppC} would offer a unique opportunity to observe this rare decay channel and test our
predictions. 

%%%%%%%%%%%%%%%%%%%%%%%%%%%%%%%%%%%%%%%%%%%%%%%%%%%
\section{Summary \label{sec-summary}}

In this work, we have carried out a comprehensive study for the rare decay of $W$ into the $B_c$ meson and a hard photon.
Since the $B_c$ meson is a heavy quarkonium, it is legitimate to use the NRQCD approach to study this process.
We have computed the NLO perturbative correction for this process in the NRQCD factorization framework.
Nevertheless, the NRQCD short-distance contains large logarithm of type $\alpha_s\ln{M_W\over M_{B_c}}$,
which may harm the convergence of perturbative expansion.
To remedy this nuisance, we also tackle this hard exclusive quarkonium production process from the angle of
the light-cone approach. The LCDA of the $B_c$ meson is further factored into the product of the short-distance coefficient function
and the NRQCD matrix element, with the former accurate to the order-$\alpha_s$.
Moreover, with the aid of the celebrated ERBL evolution equation, we are able to resum the leading collinear
logarithms $\alpha_s \ln\! {M_W\over m_Q}$ to all orders.
Finally we are able to make an optimized prediction that takes the
merits of both light-cone and NRQCD approaches.

%%%%%%%%%%%%%%%%%%%%%%%%%%%%%%%%%%%%%%%%%%%%%%%%%%%%%%%%%%%%%%%%
\begin{acknowledgments}
%%%%%%%%%%%%%%%%%%%%%%%%%%%%%%%%%%%%%%%%%%%%%%%%%%%%%%%%%%%%%%%%
We are grateful to Deshan Yang for valuable discussions on the prescription of $\gamma_5$ in the
light-cone approach.
%----------------------------
The work of F.~F. is supported by the National Natural Science Foundation of China under Grant
No.~11875318, No.~11505285, and by the Yue Qi Young Scholar Project in CUMTB.
%----------------------------
The work of Y. J. is supported in part by the National Natural Science Foundation of China
under Grants No.~11875263, No.~11475188, No.~11621131001 (CRC110 by DFG and NSFC).
%----------------------------
The work of W.-L. S. is supported by the National Natural Science Foundation of
China under Grants No.~11605144.
%----------------------------
%----------------------------
\end{acknowledgments}
%%%%%%%%%%%%%%%%%%%%%%%%%%%%%%%%%%%%%%%%%%%%%%%%%%%%%%%%%%%%%%%%

%%%%%%%%%%%%%%%%%%%%%%%%%%%%%%%%%%%%%%%%%%%%%%%%%%%

\end{document}